\documentclass[%
 reprint,
superscriptaddress,
 amsmath,amssymb,
]{revtex4-2}

\usepackage{graphicx}% Include figure files
\usepackage{dcolumn}% Align table columns on decimal point
\usepackage{bm}% bold math
\usepackage{bbold}
\DeclareMathOperator{\Tr}{Tr}
\usepackage{hyperref}% add hypertext capabilities
\usepackage[dvipsnames]{xcolor}
\usepackage{soul}
\usepackage[left]{lineno}
\usepackage{ulem}
\newcommand{\ee}{{\rm e}}

\begin{document}
\title{Topological properties of Floquet winding bands in a photonic lattice}
% 
% Topological winding metals in a photonic Floquet lattice
%Realisation of a topological winding metal in a synthetic photonic lattice

\author{Albert F. Adiyatullin}
\altaffiliation[Present address: ]
{Quandela, 7 Rue Léonard de Vinci, 91300 Massy, France}
\affiliation{Univ. Lille, CNRS, UMR 8523 -- PhLAM -- Physique des Lasers Atomes et Mol\'ecules, F-59000 Lille, France}
\author{Lavi K. Upreti}
\affiliation{Institut für Theoretische Physik und Astrophysik, Universit\"at W\"urzburg, 97074 W\"urzburg, Germany}
\author{Corentin Lechevalier}
\author{Clement~Evain}
\author{Francois~Copie}
\author{Pierre Suret}
\author{Stephane~Randoux}
\affiliation{Univ. Lille, CNRS, UMR 8523 -- PhLAM -- Physique des Lasers Atomes et Mol\'ecules, F-59000 Lille, France}
\author{Pierre Delplace}
\affiliation{ENS de Lyon, CNRS, Laboratoire de physique (UMR CNRS 5672), F-69342 Lyon, France}
\author{Alberto Amo}
\email{alberto.amo-garcia@univ-lille.fr}
\affiliation{Univ. Lille, CNRS, UMR 8523 -- PhLAM -- Physique des Lasers Atomes et Mol\'ecules, F-59000 Lille, France}

\date{\today}

\begin{abstract}
The engineering of synthetic materials characterised by more than one class of topological invariants is one of the current challenges of solid-state based and synthetic materials.
Using a synthetic photonic lattice implemented in a two-coupled ring system we engineer an anomalous Floquet metal that is gapless in the bulk and shows simultaneously two different topological properties. On the one hand, this synthetic lattice presents bands characterised by a winding number. The winding emerges from the breakup of inversion symmetry and it directly relates to the appearance of Bloch suboscillations within its bulk. On the other hand, the Floquet nature of the lattice results in well-known anomalous insulating phases with topological edge states. The combination of broken inversion symmetry and periodic time modulation studied here enrich the variety of topological phases available in lattices subject to Floquet driving and suggest the possible emergence of novel phases when periodic modulation is combined with the breakup of spatial symmetries.
%Our experiments enrich the variety of topological phases available in lattices subject to periodic time modulation, and suggest novel phases when symmetries other than the inversion symmetry are broken.
\end{abstract}

\maketitle

One of the most striking properties of topological phases of matter is the appearance of robust unidirectional interface states between two gapped materials of different topology. The existence and number of edge channels is determined by a topological invariant, which is a property of the Hamiltonian describing the bulk materials \cite{Schnyder2008, Gong2018}. This pivotal idea, known as the bulk-edge correspondence, has successfully explained the topological edge transport in the quantum Hall effect and in topological insulators \cite{Hasan2010}, the existence of topological edge states in anomalous Floquet systems~\cite{rudner_anomalous_2013} and in non-Hermitian lattices~\cite{Gong2018}. Even within the bulk, the non-trivial topology of a lattice Hamiltonian gives rise to remarkable phenomena such as the anomalous velocity due to non-zero Berry curvature \cite{Aidelsburger2014,wimmer_experimental_2017}, the quantized transport in a Thouless pump \cite{nakajima_topological_2016,lohse_thouless_2016} and the braiding of bands in non-Hermitian systems~\cite{Wang2021Nature}.

\begin{figure}[t]
\includegraphics[width=\columnwidth]{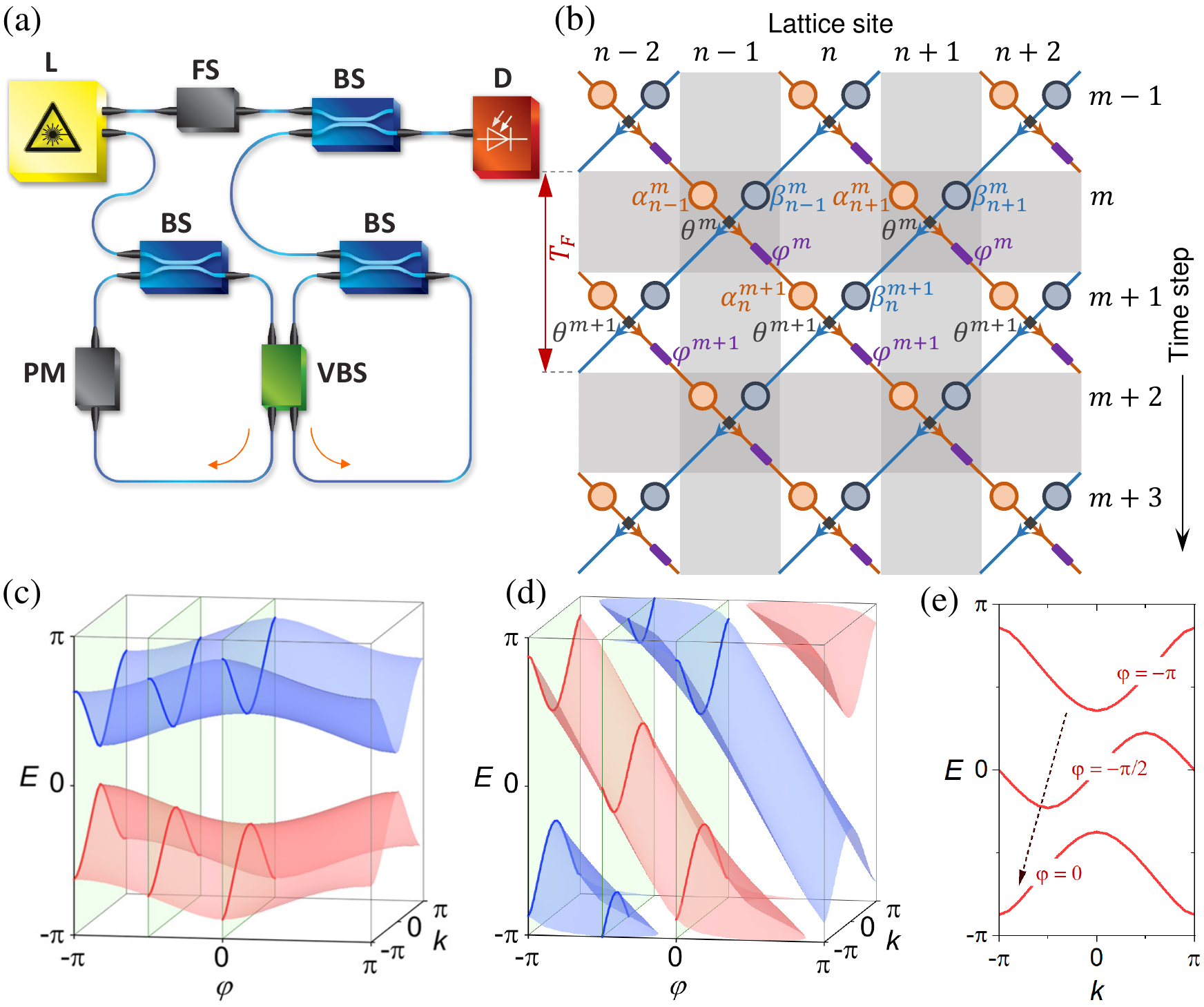}
\caption{\label{fig1} 
Floquet winding metals. 
(a) The experimental platform consists of two 40~m long fiber rings with a 0.55 m difference of length coupled via a variable beamsplitter (VBS).
One ring contains a phase modulator (PM) that controls the phase of light pulses.
(b) The dynamics in the rings can be mapped onto a lattice: propagation in the left (right) fiber ring is represented with orange (blue) lines, and lattice sites are shown with circles.
$T_F$ represents one Floquet driving period.
(c)-(d) Calculated band structure of a Floquet insulator (c) with $K=0$ ($c_1=1, c_2=-1$) and a Floquet winding metal (d) with $K=-1$ ($c_1=-2, c_2=0$). 
(e) Selected band-cuts for different values of $\varphi$ corresponding to the red band in (d). 
}
\end{figure}

Enlarging the palette of topological effects in lattices beyond the bulk-edge correspondence is an important resource that would allow combining different topological properties in a single material.
An example of band topologies with properties beyond the bulk-edge correspondance are periodically-driven (Floquet) Hamiltonians with nontrivial band holonomies. 
The eigenvalues of Floquet Hamiltonians can form bands that are periodic both in momentum and quasienergy. 
This double periodicity enables the possibility of engineering bands with non-trivial windings, that is, bands that traverse the Brillouin zone in any possible direction, even across the top and bottom of the quasienergy spectrum.
%This particular feature is one of the hallmarks of the anomalous topological edge states in finite size lattices~\cite{rudner_anomalous_2013}. 
Recently, it has been shown that when inversion symmetry is broken in a Floquet-Bloch lattice the bulk modes can also present nontrivial holonomies and windings across the Brillouin zone~\cite{Zhou2016,cedzich_complete_2018,upreti_topological_2020}. 
This situation is illustrated in Fig.~\ref{fig1}(d): two bands never touch each other, but still traverse the whole quasienergy spectrum. Since the system is gapless in the sense that bulk states exist at all energies, its spectrum can be identified with that of a metal~\cite{ying_symmetry-protected_2018}. 

Here, we report the experimental implementation of such a Floquet metal with anomalous edge states. The winding of the bulk bands, induced by a suitable inversion symmetry breaking, can be directly measured via the number of Bloch suboscillations in the dynamics of a wavepacket accelerated across the Brillouin zone. Furthermore, the time-periodic nature of the system can be used to engineer anomalous Floquet topological edge states. Therefore, Floquet-Bloch bands with broken inversion symmetry allow engineering two distinct topological properties in the same synthetic material. Thanks to a heterodyne measurement technique, we get a direct access to both the spectral bulk winding bands and to the anomalous edge states that we experimentally show to exist despite the absence of a complete gap.

To engineer these topological properties, we use a two-dimensional synthetic photonic lattice implemented in two coupled fiber rings.
Recently, photonic platforms based on fiber rings have permitted the study of unconventional topological effects hardly accessible in other systems~\cite{wimmer_experimental_2017,weidemann_topological_2020,wang_generating_2021,nitsche_eigenvalue_2019,chalabi_synthetic_2019,leefmans_topological_2022}.
The propagation of light pulses in two rings (Fig.~\ref{fig1}(a)) can be mapped into a lattice of oriented scatterers (Fig.~\ref{fig1}(b)), whose couplings and onsite energies can be manipulated at will \cite{regensburger_paritytime_2012,wimmer_optical_2013, weidemann_topological_2020,lechevalier_single-shot_2021}.
The dynamics of a light pulse injected in the system follows a split-step coherent walk described by the equations \cite{Bisianov2019,weidemann_topological_2020}:
\begin{eqnarray} \label{eq1}
\alpha_n^{m+1} &=& \left(\cos\theta_m\alpha_{n-1}^m + i\sin\theta_m\beta_{n-1}^m \right) e^{i\varphi_m}
\nonumber \\
\beta_n^{m+1} &=& i\sin\theta_m\alpha_{n+1}^m + \cos\theta_m\beta_{n+1}^m,
\end{eqnarray}
where $\alpha_n^{m}$ and $\beta_n^{m}$ denote the complex amplitudes of a light pulse in the left and right fiber ring. The temporal position of a pulse within a ring corresponds to a lattice site $n$ while the round trip number is the time step $m$.
The splitting ratio of the beamsplitters at step $m$ is parametrized by $\theta_m$ so that the reflection and transmission amplitudes are given by $\cos\theta_m$ and $\sin\theta_m$, respectively. Lastly, an electrooptical phase modulator (PM) applies an extra phase $\varphi_m$ to all light pulses in one of the rings at a time step $m$.

%**********************************
%**********************************
%**********************************
%**********************************
%**********************************
%**********************************
%**********************************

We consider a time-periodic version of the model described by Eq.~(\ref{eq1}) with two steps per period $T_F$. The coupling between rings alternates between $\theta_1$ and $\theta_2$ on odd and even steps. In the experiments and simulations presented below, we use $\theta_1 = \pi/4-0.1$ and $\theta_2 = \pi/4-0.4$.
Similarly, $\varphi_m$ takes the values $\varphi_1=c_1\varphi$ and $\varphi_2=c_2 \varphi$, where $\varphi\in[-\pi,\pi]$, and $c_{1,2}$ are integer coefficients (Fig. \ref{fig1}(b)). The periodicity of the system in synthetic space and time allows applying the Floquet-Bloch ansatz to the eigenstates of Eq.~(\ref{eq1}): $(\alpha^{m}_{n},\beta^{m}_{n})^{\dagger}=(A,B)^{\dagger}e^{-iEm/2}e^{ikn/2}$, with $E$ being the quasienergy, and $k$ the quasimomentum associated to the real-space position in the lattice.

For a fixed value of $\varphi=0$, the system has one dimension and it presents anomalous edge modes for specific values of the splitting ratios $\theta_1$ and $\theta_2$, as studied in Ref.~\cite{Bisianov2019}.
Interestingly, the phase $\varphi$ can be seen as an additional parametric dimension, with periodicity between $[-\pi, \pi]$. In this way the model becomes two-dimensional with two bands $E_{\pm}(k,\varphi)$ defined in the generalised momenta space defined by $k$ and the parametric dimension $\varphi$ (see Fig.~\ref{fig1}(c),(d)) \cite{upreti_topological_2020,suppl}.
The use of parametric dimensions has been very successful in augmenting the available dimensions in synthetic materials and in exploring topological order in quasi-crystals~\cite{kraus2012topological, Baboux2017}, Berry curvature in photonic bands~\cite{wimmer_experimental_2017}, the four-dimensional quantum Hall effect~\cite{Lohse2018, Zilberberg2018} and nonlinear Thouless pumping~\cite{Jurgensen2021}.

The periodicity of the Brillouin zone in $k$, $\varphi$ and $E$ allows for the engineering of bands with nontrivial windings.  
An example of such peculiar band structure is shown in Fig.~\ref{fig1}(d).
The bands are inclined in quasienergy: when $\varphi$ is changed, they experience a shift in quasienergy and a lateral displacement along quasimomentum $k$ (Fig.~\ref{fig1}(e)), the combined effect resulting in their winding.

Insights into the topological character of the winding of the bands can be gained by looking at the evolution operator after one Floquet period (two steps in our model):
\begin{equation} \label{eq2}
U_F(k,\varphi) = e^{iK\varphi} T_2 S_2 T_1 S_1,
\end{equation}
where the unitary operators $S_{1,2}$ and $T_{1,2}$ represent the action of beamsplitters and phase shifts along the lattice, and $K\equiv (c_1+c_2)/2$ \cite{suppl}. 
%The non-zero matrix elements of $U_F$ in the real-space picture are sketched in Fig. \ref{fig1}(c).
From Eq.~\eqref{eq2} one can see that $K\neq 0$ imprints an additional net phase to one of the rings during one Floquet period and breaks the generalized inversion symmetry $U_F(k,\varphi) \leftrightarrow U_F(-k,-\varphi)$, leading to the winding of the bands along the quasienergy direction. This net phase gained by light travelling in the left ring can be seen as an onsite potential (see~\cite{suppl} for a Hamiltonian description of the model) and it cannot be gauged away. The phase added periodically by the modulator is not a trivial shift of the model and, as we will see in the following, has strong consequences in the dynamics of wavepackets.

%**********************************
%**********************************
%**********************************
\begin{figure}[t]
\includegraphics[width=\columnwidth]{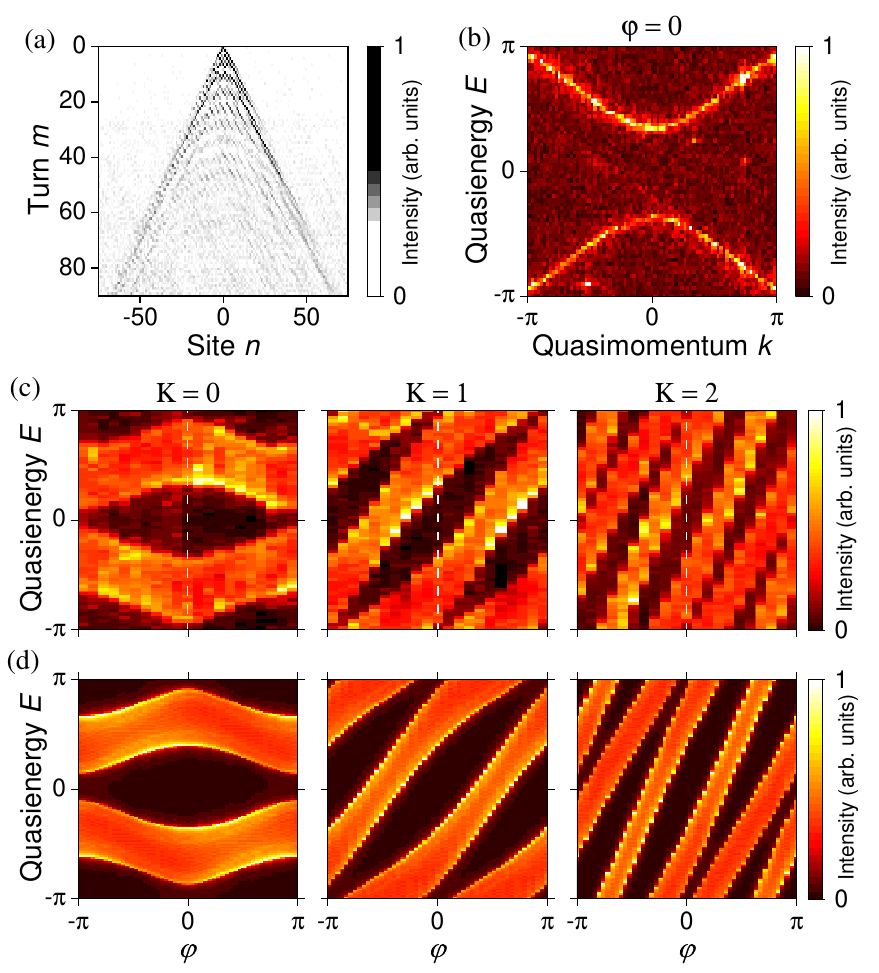}
\caption{\label{fig2} 
Tomography of the quasienergy bands. 
(a) Experimentally observed split-step coherent walk for $\varphi=0$.
(b) Reconstructed band structure of the system. 
(c) Measured band tomographies for $K=0$ ($c_1=1,c_2=-1$), $K=1$ ($c_1=2,c_2=0$), and $K=2$ ($c_1=3,c_2=1$) integrated over the quasimomentum $k$.
(d) Band structure from simulations of Eq.~\ref{eq1} for the same parameters as in (c).
}
\end{figure}

%**********************************
%**********************************
%**********************************

%**********************************
%**********************************
%**********************************

\begin{figure}[t!]
\includegraphics[width=\columnwidth]{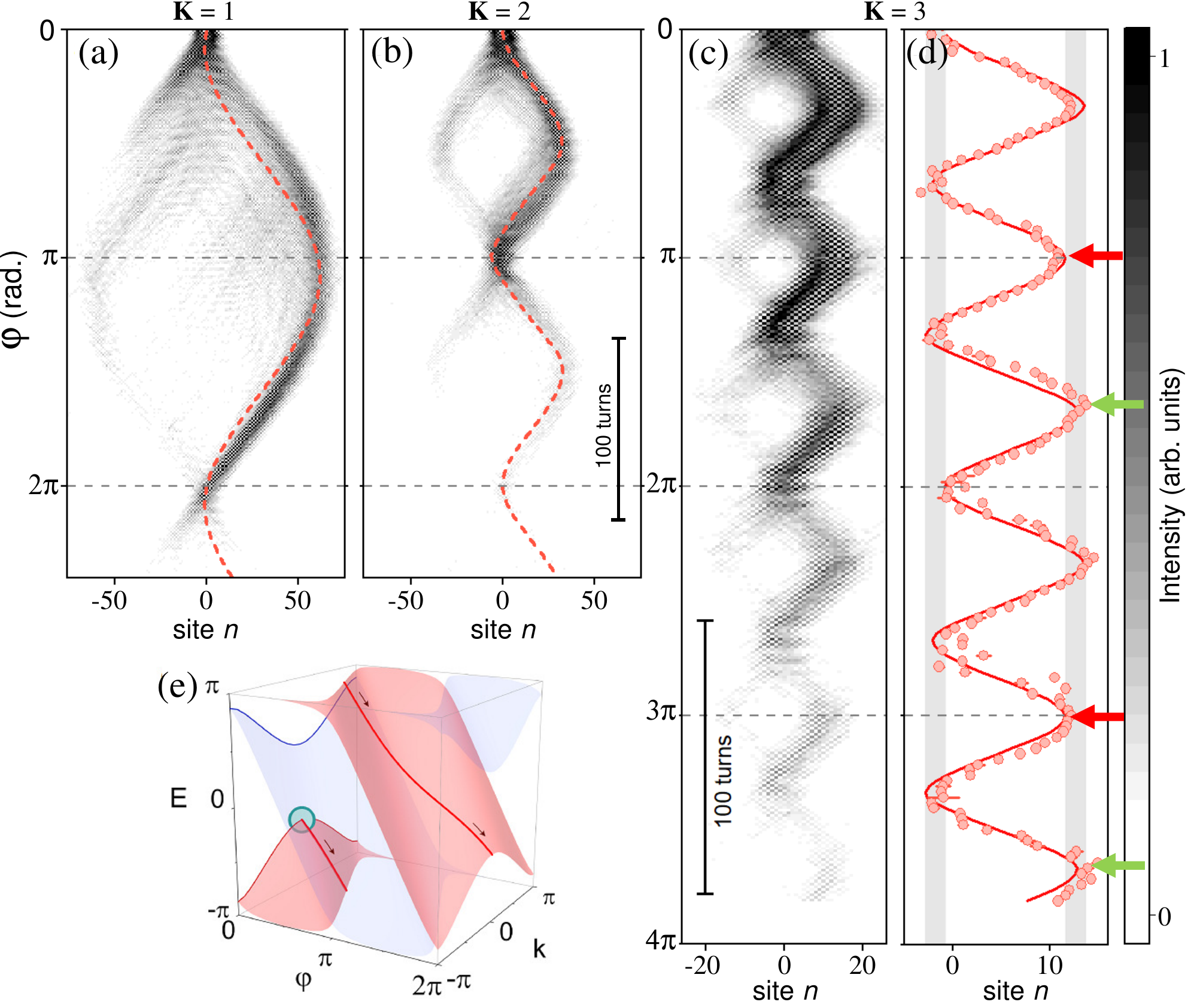}
\caption{\label{fig3} 
Topological Bloch suboscillations.
(a)-(c) Measured real-space evolution of a wavepacket injected close to $k=0$ into one of the bands and evolved under an adiabatic increase of $\varphi$ for (a) $K=1$ ($c_1=2,c_2=0$), (b) $K=2$ ($c_1=5,c_2=-1$), and (c) $K=3$ ($c_1=8,c_2=-2$). Note that in (a) and (b), the initial wavepacket has a momentum slightly smaller and larger, respectively, than $k=0$ due to the experimental injection technique.
Dashed orange lines show analytical curves. The increase rate $d\varphi/dt$ is $2\pi\cdot0.008$ rad/turn for (a) and (b), and $2\pi\cdot0.012$ rad/turn for (c).
(d) Dots: Measured evolution of the center-of-mass of the wavepacket in (c); solid line: analytic solution.
Error bars represent $1\sigma$ confidence intervals and generally are smaller than the dot size.
Gray areas emphasize the difference between the maximal and the minimal achievable amplitudes of suboscillations in the analytic curve.
(e) Illustration of the experimental procedure.
}
\end{figure}

%**********************************
%**********************************
%**********************************

The quasienergy winding is a topologically protected property of the bulk of the system. The corresponding invariant can be defined using a homotopic property of $U_F$ \cite{kitagawa_topological_2010}:
\begin{equation}\label{eq:win}
\nu = \sum_{j=\pm} \frac{1}{2\pi }\int_{0}^{2\pi}  d\varphi \frac{\partial E_j}{\partial \varphi}
= \frac{1}{2\pi i}\int_{0}^{2\pi}  d\varphi \Tr \left[U_{F}^{-1}\frac{\partial U_{F}}{\partial\varphi}\right],
\end{equation}
which gives $\nu=2K$ \cite{suppl}. Since our model features two bands, the number $K$ has a simple meaning: it shows how many times each band winds along the quasienergy axis for one full turn of $\varphi$ from $-\pi$ to $\pi$. 
Note that Eq.~\eqref{eq:win} does not depend of $k$: the winding is a property of the $\varphi$ synthetic dimension and it takes the same value for any value of $k$.

We experimentally demonstrate the Floquet winding metals by injecting a single $\approx1$~ns long laser pulse at a position $\alpha_{n=0}^{m=1}$ and following the dynamics of the system at each time step (Fig.~\ref{fig2}(a)). Such localised excitation populates all the bands of the model.
We get access to both the amplitude and the phase of light $\alpha_n^m,\beta_n^m$ in each ring at each lattice site $n$ and time step $m$ by using an optical heterodyning technique  \cite{lechevalier_single-shot_2021}.
For this, we let the light pulse at each position and time step interfere with a local oscillator shifted by 3 GHz from the frequency of the laser used to inject the initial pulse. By Fourier transforming the beating of the signal and the local oscillator we can directly reconstruct the bands as shown in Fig. \ref{fig2}(b). See Ref.~\cite{suppl} for further details.

For $\varphi=0$ (Fig. \ref{fig2}(b)) the system features two symmetric bands with respect to $E=0$. We repeat such measurement for values of $\varphi$ from $-\pi$ to $\pi$, thus performing a full tomography of the band structure $E(k,\varphi)$.
The measured tomographies integrated along the quasimomentum direction, for  $c_1$ and $c_2$ corresponding to $K=0$, $1$, and $2$ are presented in Fig.\ref{fig2}(c). 
All subplots feature two distinct bands, each of which wraps $K$ times along the quasienergy axis.
These results are in perfect agreement with numerical simulations in Fig.\ref{fig2}(d).

The topological feature we have just described does not present any particular effect on the real-space edges of the lattice. 
However, it has direct consequences on the wavepacket dynamics of the system: it manifest itself in a new kind of Bloch suboscillations\cite{upreti_topological_2020}. If we adiabatically propagate a wavepacket with quasimomentum $k$ along the $\varphi$ dimension as sketched in Fig.~\ref{fig3}(e), the group velocity $v_g = \partial E(k,\varphi)/\partial k$ periodically changes its sign, resulting in suboscillations of the wavepacket. 
Analytical inspection of the expression for $v_g$ shows that within one Bloch period the group velocity changes its sign $2K$ times, thus leading to observation of $K$ suboscillations. 
The number of suboscillations is thus determined by the winding number. It is independent of the coupling parameters $\theta_1$, $\theta_2$, and it is preserved in the presence of a weak spatial disorder in the couplings (see Refs.~\cite{upreti_topological_2020,suppl} for further details). This feature is present as long as the wavepacket dynamics is adiabatic and out of the particular case when the bands are flat (i.e., $\theta_1$, $\theta_2 = \pi/2$), for which there are no Bloch oscillations at all.

To observe the topological suboscillations, we prepare a wavepacket at $k\approx0$ in one of the quasienergy bands~\cite{suppl} and follow its evolution while $\varphi$, imprinted by the phase modulator, is adiabatically increased at a constant rate $\partial\varphi/\partial t$.
The observed dynamics for winding metals with $K=1$, $2$, and $3$ is shown in Fig. \ref{fig3}(a)-(c) respectively. The wavepacket shows an oscillatory behavior towards positive values of the lattice sites. The weak signal in the other direction arises from residual initial excitation of the other band.
While the full period of oscillations is always equal to $\Delta \varphi=2\pi$, there are exactly $K$ suboscillations over one full period. 
An analytical calculation of the wavepacket trajectory, shown in Fig. \ref{fig3}(a)-(c) with dashed lines, reproduces the observed behavior.

When $K=1$, the system shows a single oscillation over a full period $\Delta \varphi=2\pi$. This matches the expected behaviour for the usual Bloch oscillations of a wave packet accelerated by an electric field over the Brillouin zone. Indeed, in this case, there is a gauge transformation that links the dynamics under an adiabatic increase of $\varphi$ to the dynamics of a wave packet in a lattice subject to a static potential gradient (i.e., a constant electric field), as discussed in Refs.~\cite{Wimmer2015, upreti_topological_2020}. For higher values of $K$, suboscillations appear within a period of acceleration ($\varphi \rightarrow \varphi+2\pi$).
%The winding number $\nu$ topologically protects the number of suboscillations within a period (see Methods).
Interestingly, in general, the suboscillations do not have a constant amplitude. % a behaviour that departs from the trivial Bloch oscillations. 
Figure~\ref{fig3}(d) shows evidence of the variations of amplitude within a Bloch period (compare green and red arrows) for a wavepacket adiabatically accelerated in a lattice with $K=3$ over two periods of $\Delta \varphi$. 
%The position of the center-of-mass of the wavepacket is displayed in Fig.~\ref{fig3}f with dots.
%During one Bloch period ($\varphi \rightarrow \varphi+2\pi$), different suboscillations have different amplitudes. 
%, however, the overall periodicity $\varphi \rightarrow \varphi+2\pi$ is preserved.
These amplitude variations allow identifying in an unambigous manner the overall period of the Bloch oscillations, and show that the appearance of suboscillations cannot be explained by a redefinition of the periodicity of the dynamics. The observed behavior matches well the analytical calculations (solid line in Fig.~\ref{fig3}(d)).

%****************************************
%****************************************
%****************************************
\begin{figure}[t!]
\includegraphics[width=\columnwidth]{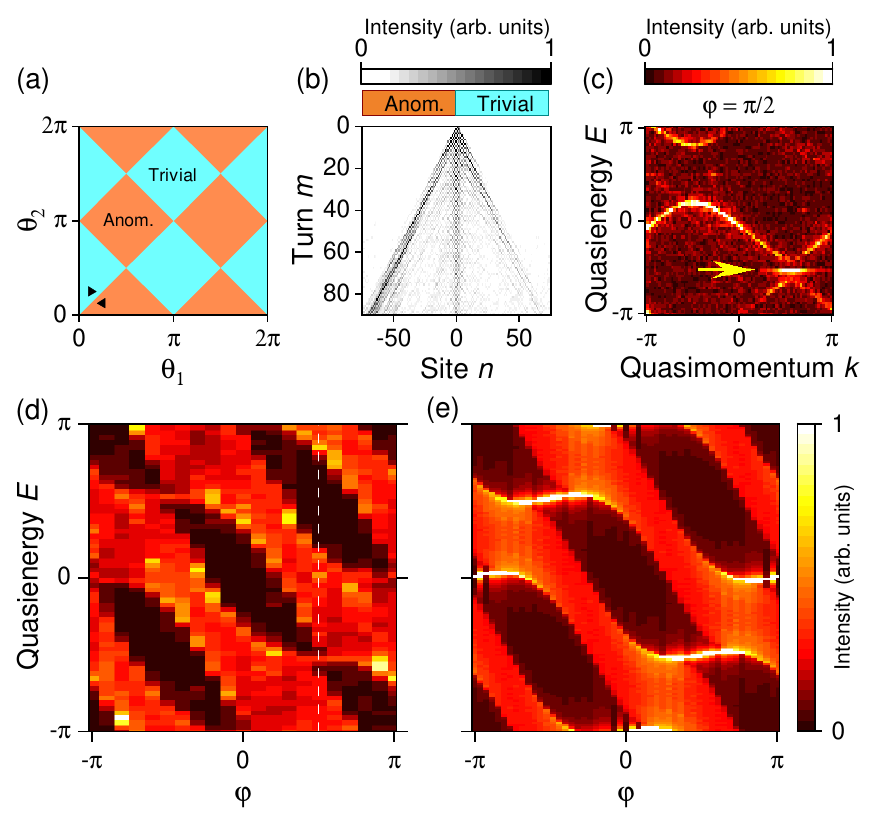}
\caption{\label{fig4} 
Topological edge states.
(a) Phase diagram of anomalous Floquet phases as a function of the coupling amplitudes in the first~$\theta_1$ and second step~$\theta_2$.
(b) Measured dynamics when exciting the lattice at a single site located at the interface between two lattices belonging to two different phases (triangles in (a)), showing a localized interface state. Both lattices are prepared with $K=-1$ ($c_1=1,c_2=-3$) and $\varphi=\pi /2$ but different values of $\theta_1$ and $\theta_2$. (c) Measured dispersion showing a flat band in one gap (shown with the arrow), corresponding to the interface state.
(d) Experimental and (e) theoretical band tomography of the winding metal for all values of $\varphi$ confirming the presence of edge state bands in each of the gaps.
}
\end{figure}
%******************************************
%****************************************
%****************************************

Finally, we demonstrate that Floquet winding metals can support a second topological property: the emergence of anomalous chiral edge states.
They arise neither from the winding number $\nu$ nor from the Chern number that vanishes due to the phase rotation symmetry~\cite{delplace_phase_2017}. They rather emerge from the generalised Floquet topological invariant related to the micromotion of the system during one driving period~\cite{rudner_anomalous_2013}. 
Such anomalous Floquet phases have been reported in 1D photonic lattices \cite{Kitagawa2012,cardano_detection_2017,bellec_non-diffracting_2017, Bisianov2019} and in 2D systems \cite{maczewsky_observation_2017,mukherjee_experimental_2017, wintersperger_realization_2020, zhang_superior_2021}. Here we show spectral evidence of the anomalous topological edge states and that they can also exist in a Floquet winding metal.

The phase diagram for the anomalous Floquet phases in the topological system is determined by the values of $\theta_1$ and $\theta_2$ for which the gap between the two bands closes (Fig.~\ref{fig4}(a)), and it does not depend on the winding $K$. Following Ref.~\cite{delplace_phase_2017}, a bulk topological invariant can be constructed to account for the number of edge states in the anomalous regime for $K=0$ (orange areas in the figure). These anomalous phases are preserved for any value of $K$ as confirmed by simulations via the presence of edge states at the edge of a single semi-infinite lattice~\cite{suppl}.

In our experiment we take profit of the full control over the couplings between the lattice sites to engineer interfaces between different anomalous topological phases.
To demonstrate this, we consider a winding metal with $K=-1$ and prepare two topologically different phases with an interface at  position $n=0$.
For lattice sites $n<0$ we set $\theta_1=\pi/4$, $\theta_2=\pi/4-0.4$, forming an anomalous phase. For $n>0$ we create a trivial phase with $\theta_1=\pi/4-0.4$, $\theta_2=\pi/4$ (triangles in Fig.~\ref{fig4}(a)).
When exciting the interface with a single pulse, the system shows a localized edge state at $n=0$ (see Fig. \ref{fig4}(b) for $\varphi=\pi/2$). Simultaneously, the band structure reveals a flat band in one of the gaps (Fig. \ref{fig4}(c)), which can be associated to the localised interface state.
To probe the full dispersion of the edge states in $k$ and $\varphi$ we perform the full band tomography (Fig.~\ref{fig4}(d)).
The characteristic spectral flow of edge states between two bands is evident in both gaps, in good agreement with simulations (Fig.~\ref{fig4}(e)). Remarkably, edge states are present even in the absence of a complete gap. While the topological origin of the edge states is confirmed by the fact that it requires the presence of an interface between two different phases, the access to the topological invariant associated to this two-dimensional split-step Floquet operator and the robustness against scattering to bulk modes in the gapless phases is an interesting question to be addressed in subsequent works.

We have shown the experimental realisation of an anomalous Floquet metal, which simultaneously hosts two different topological properties. Whereas the first one appears as a consequence of the breakup of inversion symmetry and manifests in Bloch suboscillations, the second one leads to the formation of edge states. Both of these topological properties arise from the Floquet nature of the system and therefore do not have static counterparts. The flexibility of our platform paves the road to studies of Floquet winding bands with unconventional dispersion in higher dimensions, and open unprecedented perspectives in the search for novel Floquet topological phases when combined with selected spatial symmetries or when including, for instance, non-Hermitian hoppings~\cite{Wang2021Nature,wang_generating_2021,weidemann_topological_2020}.

\textit{Acknowledgments}. We thank S.~Ravets and J.~Bloch for fruitful discussions. This work was supported by European Research Council grant EmergenTopo (865151), the H2020-FETFLAG project PhoQus (820392), the QUANTERA project Interpol (ANR-QUAN-0003-05), the French government through the Programme Investissement d’Avenir (I-SITE ULNE / ANR-16-IDEX-0004 ULNE) managed by the Agence Nationale de la Recherche, the Labex CEMPI (ANR-11-LABX-0007) and the CPER Photonics for Society P4S. LKU acknowledges support from the Deutsche Forschungsgemeinschaft (DFG, German Research Foundation) through Project-ID 258499086 - SFB 1170 and through the W\"urzburg-Dresden Cluster of Excellence on Complexity and Topology in Quantum Matter -- \textit{ct.qmat} Project-ID 39085490 - EXC 2147.

\bibliography{biblio}% Produces the bibliography via BibTeX.

%%%%%%%%%%%%%%%%%%%% Merge with supplemental materials %%%%%%%%%%%%%%%%%%%%%%%%
\widetext
\newpage
\begin{center}
\textbf{\large Supplemental Materials for: Multi-topological Floquet metals in a photonic lattice}
\end{center}
%%%%%%%%%% Merge with supplemental materials %%%%%%%%%%
%%%%%%%%%% Prefix a "S" to all equations, figures, tables and reset the counter %%%%%%%%%%
\setcounter{equation}{0}
\setcounter{figure}{0}
\setcounter{table}{0}
\setcounter{page}{1}
\makeatletter
\renewcommand{\theequation}{S\arabic{equation}}
\renewcommand{\thefigure}{S\arabic{figure}}
% \renewcommand{\bibnumfmt}[1]{[S#1]}
% \renewcommand{\citenumfont}[1]{S#1}

%*************************************************************
%*************************************************************
\section{Calculation of the band structure}
By applying the Floquet-Bloch ansatz to Eq.~(1) in the main text and solving the determinant problem we obtain the solution for the energies of the bands $E_{\pm}$:

\begin{eqnarray}\label{eq:EE}
E_{\pm}(k,\varphi) =\pm\arccos[\cos\theta_{1}\cos\theta_{2}\cos\left(k+K\varphi\right) \nonumber \\
- \sin\theta_{1}\sin\theta_{2}\cos\left(\Delta\varphi\right)] + K\varphi,\nonumber
\end{eqnarray}

\noindent where $\varphi_{1,2}=c_{1,2}\varphi$, $K\equiv(c_1+c_2)/2$, $\Delta\equiv(c_1-c_2)/2$.
In this work we consider the case of integer $K$ and $\Delta$, which makes the period along the $\varphi$ direction equal to $2\pi$.
The last term $K\varphi$ emphasizes the fact that each band winds $K$ times along the quasienergy axis when $\varphi$ is changed by $2\pi$.

\section{Derivation of the Floquet evolution operator \label{Sec:Floquet}}

For the Floquet period of 2 steps the evolution of the system in real space can be written as
\begin{equation}
\Psi(m+2) = U\Psi(m),
\end{equation}
where
\begin{equation}
\Psi(m) = 
\begin{pmatrix}
\cdots \\
\alpha_{n}^m \\
\beta_{n}^m \\
\alpha_{n+2}^m \\
\beta_{n+2}^m \\
\cdots
\end{pmatrix}
\end{equation}
is a vector representing the state of the system in real space at time step $m$, and
\begin{equation}
U = \sum_{x_i,y_j}U_{x_i\rightarrow y_j}\left|y_j\right>\left<x_i\right|
\end{equation}
is the real-space Floquet evolution operator.
Here $\left|x_i\right>$ and $\left|y_j\right>$, where $x,y\in\{\alpha,\beta\}$ and $i,j$ are the site number, represent a vector $\Psi$ with $x_i=1$ (or $y_j=1$) and all the other components equal to zero.
%($x_i$ and $y_j$ denote the lattice sites).
Non-zero matrix elements of $U$ can be found from the evolution equation (Eq.~(1) of the main text):
\begin{align*}
U_{\alpha_{n}\rightarrow\beta_{n}} &= is_1Re^{i\Delta\varphi} & U_{\beta_{n}\rightarrow\alpha_{n}} &= is_1Re^{-i\Delta\varphi} \nonumber\\
U_{\beta_{n}\rightarrow\alpha_{n+2}} &= is_2Re^{iK\varphi} &
U_{\alpha_{n}\rightarrow\beta_{n-2}} &= is_2Re^{-iK\varphi} \nonumber\\
U_{\alpha_{n}\rightarrow\alpha_{n+2}} &= s_3Re^{iK\varphi} &
U_{\beta_{n}\rightarrow\beta_{n-2}} &= s_3Re^{-iK\varphi} \nonumber\\
U_{\beta_{n}\rightarrow\beta_{n}} &= -s_4Re^{i\Delta\varphi} &
U_{\alpha_{n}\rightarrow\alpha_{n}} &= -s_4Re^{-i\Delta\varphi} 
\end{align*}
where $s_1=\cos\theta_1\sin\theta_2$, $s_2=\sin\theta_1\cos\theta_2$, $s_3=\cos\theta_1\cos\theta_2$, $s_4=\sin\theta_1\sin\theta_2$, and $R=e^{iK\varphi}$.
%These couplings are schematically shown in Fig. 1C of the main text.
\\

To obtain the Floquet evolution operator in reciprocal space, we can use the Floquet-Bloch ansatz and substitute it into the evolution equation. This gives
\begin{equation}\label{eq:floquet}
U_F(k,\varphi) = 
\begin{pmatrix}
e^{i(\varphi_1+\varphi_2)}e^{-ik}\cos\theta_1\cos\theta_2 &-& e^{i\varphi_2}\sin\theta_1\sin\theta_2
&
ie^{i(\varphi_1+\varphi_2)}e^{-ik}\sin\theta_1\cos\theta_2 &+& ie^{i\varphi_2}\cos\theta_1\sin\theta_2
\\
ie^{ik}\sin\theta_1\cos\theta_2 &+& ie^{i\varphi_1}\cos\theta_1\sin\theta_2
&
e^{ik}\cos\theta_1\cos\theta_2 &-& e^{i\varphi_1}\sin\theta_1\sin\theta_2
\end{pmatrix}.
\end{equation}\\
It can be seen that the Floquet evolution operator can be factorized in a sequential manner
\begin{equation}\label{eq:flfact}
U_F(k,\varphi) = D_2B_2(k)S_2D_1B_1(k)S_1
\end{equation}
where $S_{1,2} = S(\theta_{1,2})$ are scattering matrices representing the action of the beamsplitter,
\begin{equation}\label{eq:S}
S(\theta) = 
\begin{pmatrix}
\cos\theta & i\sin\theta \\
i\sin\theta & \cos\theta
\end{pmatrix},
\end{equation}
$B_{1,2}(k)$ are translation operators
\begin{equation}\label{eq:B}
B_1(k) = 
\begin{pmatrix}
1 & 0 \\
0 & e^{ik}
\end{pmatrix},
B_2(k) = 
\begin{pmatrix}
e^{-ik} & 0 \\
0 & 1
\end{pmatrix},
\end{equation}
and $D_{1,2}$ correspond to the phase shift on odd and even steps:
\begin{equation}\label{eq:D}
D_{1,2} = 
\begin{pmatrix}
e^{i\varphi_{1,2}} & 0 \\
0 & 1
\end{pmatrix}.
\end{equation}\\

To study the symmetry properties of the unitary evolution operator we symmetrize the matrices $B_{1,2}(k)$ and $D_{1,2}$:
\begin{equation}
B(k) \equiv 
\begin{pmatrix}
e^{-ik/2} & 0 \\
0 & e^{ik/2}
\end{pmatrix},
D(\varphi) \equiv
\begin{pmatrix}
e^{i\varphi/2} & 0 \\
0 & e^{-i\varphi/2}
\end{pmatrix}
\end{equation}
and write 
\begin{eqnarray}
U_F(k,\varphi) 
&=& e^{i(\varphi_1+\varphi_2)/2} D(\varphi_2) B(k) S_2 D(\varphi_1) B(k) S_1 \nonumber\\
&=& e^{i(\varphi_1+\varphi_2)/2} T_2 S_2 T_1 S_1,
\end{eqnarray}
where $T_{1,2} =  D(\varphi_{1,2}) B(k)$.
We notice that both $B(k)$ and $D(\varphi)$ possess inversion symmetry: $\sigma_xB(k)\sigma_x=B(-k)$, $\sigma_xD(\varphi)\sigma_x=D(-\varphi)$, where $\sigma_x$ is the Pauli matrix. 
Consequently, for $\varphi_1+\varphi_2=0$ the Floquet evolution operator also has the inversion symmetry $\sigma_x  U_F(k,\varphi)\sigma_x=U_F(-k,-\varphi)$. 
However, introducing a net phase $\varphi_1+\varphi_2\neq0$ over one Floquet period breaks this symmetry and leads to winding of the bands.

%*************************************************************
%*************************************************************
\section{Calculation of the topological invariant}

Given the factorized version of the Floquet evolution operator \eqref{eq:flfact}, we can calculate the topological invariant
\begin{eqnarray}
\nu &=& \frac{1}{2\pi i}\int_{0}^{2\pi}  d\varphi \Tr \left[U_{F}^{-1}\frac{\partial U_{F}}{\partial\varphi}\right] \nonumber\\
&=& \frac{1}{2\pi i}\int_{0}^{2\pi}  d\varphi \Tr \left[S_1^\dagger B_1^\dagger D_1^\dagger S_2^\dagger B_2^\dagger D_2^\dagger \frac{\partial} {\partial\varphi}\left[D_2B_2S_2D_1B_1S_1\right]\right] \nonumber\\
&=& \frac{1}{2\pi i}\int_{0}^{2\pi}  d\varphi \Tr \left[D_1^\dagger\frac{\partial D_1} {\partial\varphi} + D_2^\dagger\frac{\partial D_2} {\partial\varphi}\right].
\end{eqnarray}
By substituting \eqref{eq:D} we get
\begin{equation}
\nu = \frac{1}{2\pi i}\int_{0}^{2\pi}  d\varphi \left[ic_1+ic_2\right] = 2K.
\end{equation}

%*************************************************************
%*************************************************************
\section{Hamiltonian formalism }
The coherent split step model discussed above in terms of evolution operators can also be described using a time dependent Hamiltonian, as originally discussed in Ref.~\cite{upreti_topological_2020}. To describe the two step process, we divide each period of the Hamiltonian evolution into four steps:

\begin{align}
H(t,k_x) =
 \left\{
\begin{array}{ll}
H_1(k_x) = \begin{pmatrix}
0 & 0 \\
0  & -V_{1}
\end{pmatrix}, \quad &0<t\le t_1\\
H_2(k_x) =  \begin{pmatrix}
0 & -J_{1}  \text{e}^{-i k_{x}/2}\\
-J_{1}  \text{e}^{i k_{x}/2} & 0
\end{pmatrix}\quad &t_1<t\le t_2\\
H_3(k_x) = \begin{pmatrix}
0 & 0 \\
0  & -V_{2}
\end{pmatrix} \quad &t_2<t\le t_3\\
H_4(k_x) = \begin{pmatrix}
0 & -J_{2} \text{e}^{i k_{x}/2} \\
-J_{2} \text{e}^{-i k_{x}/2}  & 0
\end{pmatrix}  \quad  &t_3<t\le T
\end{array}
\right.
\end{align}

\noindent The parameters $V_1, V_2, J_1$ and $J_2$ are related to the phases gained in the left ring and the coupler strengths via:

\begin{align}
 \phi_{1} \equiv V_{1} \tau_1/\hbar \quad \quad \theta_{1} \equiv J_{1}\tau_2/\hbar  \quad \phi_{2} \equiv V_{2} \tau_3/\hbar \quad \theta_{2} \equiv J_{2}\tau_4/\hbar .
\end{align}

The action of the phase modulator in the first and third steps directly maps into a modification of the onsite energy, while the couplers act as hopping amplitudes. In this way it can be readily seen that the presence of the phase modulation breaks inversion symmetry.

Note that the each time step can be associated to an evolution operator $U_j$ during the duration $\tau_j=t_j-t_{j-1}$:

 \begin{align}\label{eq:wind_hamiltonian2}
U_{j}(k_x) \equiv \text{e}^{-i H_{j}(k_x) \tau_j/\hbar},
\end{align}

\noindent so that the evolution (Floquet) operator after one full period is defined as  $ U_{F}(k_x) = U_{4} U_{3} U_{2} U_{1} $, with stepwise evolution operators:

\begin{align}\label{eq:U1matrix}
U_1=  \begin{pmatrix}
1 & 0 \\
0  & \text{e}^{i \phi_{1}}
\end{pmatrix}
\qquad U_2=   \begin{pmatrix}
\cos \theta_{1} & i \ee^{-i k_{x}/2}\sin \theta_{1} \\
i \ee^{i k_{x}/2}\sin \theta_{1}  & \cos \theta_{1}
\end{pmatrix}\\
U_3=  \begin{pmatrix}
1 & 0 \\
0  & \text{e}^{i \phi_{2}}
\end{pmatrix}
\qquad U_4=   \begin{pmatrix}
\cos \theta_{2} & i \ee^{-i k_{x}/2}\sin \theta_{2} \\
i \ee^{i k_{x}/2}\sin \theta_{2}  & \cos \theta_{2},
\end{pmatrix}
\end{align}
\noindent leading to the expression of the Floquet operator described in Sec.~~\ref{Sec:Floquet}.

%*************************************************************
%*************************************************************
\section{Topologically protected Bloch sub-oscillations}
The group velocity in the real space dimension can be found as
% \begin{eqnarray}\label{eq:bloch}
\begin{equation}
v_{g}^{\pm}(k,\varphi) = \dfrac{\partial E_\pm(k,\varphi)}{\partial k} \nonumber \\
= \frac{\pm\cos \theta _1 \cos \theta _2 \sin \left(k+K\varphi\right)}{\sqrt{1-\left[\cos \theta _1 \cos \theta _2 \cos \left(k+K\varphi\right)-\sin \theta _1 \sin \theta _2 \cos \left(\Delta\varphi\right)\right]^2}} \nonumber
\end{equation}
% \end{eqnarray}
Due to the term $\sin \left(k+K\varphi\right)$ in the numerator, the sign of $v_{g}$ changes $2K$ times when $\varphi$ is changed by $2\pi$, forcing a wavepacket to experience $K$ sub-oscillations during one driving period.
Since $\sin \left(k+K\varphi\right)$ becomes zero with periodicity of $\pi/K$ in $\varphi$, we can claim that the winding number topologically protects the frequency of Bloch sub-oscillations.
At the same time if $K\neq\Delta$, then the term $\cos \left(k+K\varphi\right)$ in the denominator precesses at a different rate than $\cos \left(\Delta\varphi\right)$. 
Consequently, the translational symmetry $v_g(k,\varphi)=v_g(k,\varphi+2\pi/K)$ gets broken, leading to sub-oscillations of different amplitudes.
\\

Finally, the center-of-mass motion of the wavepacket can be found by integrating the group velocity:
\begin{equation}\label{eq:com}
X(k,t) = \int_{0}^{t} v_g(k,\varphi(\tau)) \,d\tau
\end{equation}

%*************************************************************
%*************************************************************
\section{Experimental platform}
The photonic network is made of two fiber rings coupled by an electronically-controlled high-bandwidth variable beamsplitter (EOSpace AX-2x2-0MSS-20).
Each of the rings contains an erbium-doped fiber amplifier (Keopsys CEFA-C-HG) followed by a narrow-band optical filter (EXFO XTM-50), an isolator, a polarizer, a variable attenuator, and an optical switch (Photonwares NSSW). 
One of the rings contains a phase electro-optic modulator (EOM, iXblue MPZ-LN-10), which imposes the phases $\varphi_{1,2}$.
All the fiber components use polarization-maintaining fibers.
Each ring has a length of 40 m, and the length difference between the rings is 0.55 m.
The mean length of the two rings sets the round trip period, of 205 ns, between the different time steps $m$. The length difference sets the temporal size of the lattice sites $n$ in the synthetic spatial dimension, of 2.7 ns.
\\

For the injection of light, emission of a narrow single-frequency laser (IPG Photonics ELR-5-LP) at a wavelength of 1550 nm is chopped into 1.4 ns-long pulses by an amplitude EOM (iXblue MXER-LN-10).
Before entering the fiber rings the light passes through an optical switch, which is closed after the injection. 
This ensures that no spurious signal from the laser enters the fibers during the experiment.
The prepared injection signal is coupled into one of the rings through a 70/30 beamsplitter.
\\
%*************************************************************
%*************************************************************
\section{Measurement procedure}
The light field in the system is probed via an 80/20 beamsplitter in each of the rings.
To get access to both the amplitude and the phase of each light pulse we use optical heterodyning.
For this, a fraction of the laser light is modulated by a phase EOM at a frequency of $\Omega=3$ GHz, thus creating sidebands shifted by $\pm\Omega$ from the laser frequency.
Next, the $+\Omega$ sideband is filtered out by a home-built fiber ring cavity actively locked to this sideband.
The filtered out light field is used as a local oscillator, and its beating with the signal from each ring is measured by a fast photodiode (Thorlabs DET08CFC, 5 GHz).
Recording the response of the photodiode with a fast oscilloscope (Tektronix MSO64, bandwidth 4 GHz) allows to see the beating, the amplitude and the phase of which directly correspond to the amplitude and the phase of the light field under study. 
By reshaping the measured signal in a two-dimensional matrix, we can observe the coherent walk in the real space (Fig.~2(a) of the main text).
Performing a two-dimensional Fourier transform of the coherent walk gives access to the band structure centered at the frequency $\Omega$ of the local oscillator (Fig.~2(b) of the main text).
\\

\section{Reconstruction of the band structure}

Due to the periodicity of the system in both synthetic dimension and time, its band structure can be obtained simply by calculating the two-dimensional Fourier transform (2DFT) of a split-step coherent walk.
An important prerequisite for this is that each site of the walk ($\alpha_n^m$ and $\beta_n^m$) is a complex number, which accounts for both the amplitude and the phase of the light field.
In our experiment the measured quantity is the beating of the signal with the local oscillator at a constant frequency $\Omega$.
This allows us to reconstruct the band structure by performing the 2DFT of the measured signal and offsetting it by the frequency $\Omega$. 
\\

\begin{figure}[h]
\includegraphics[width=17.2cm]{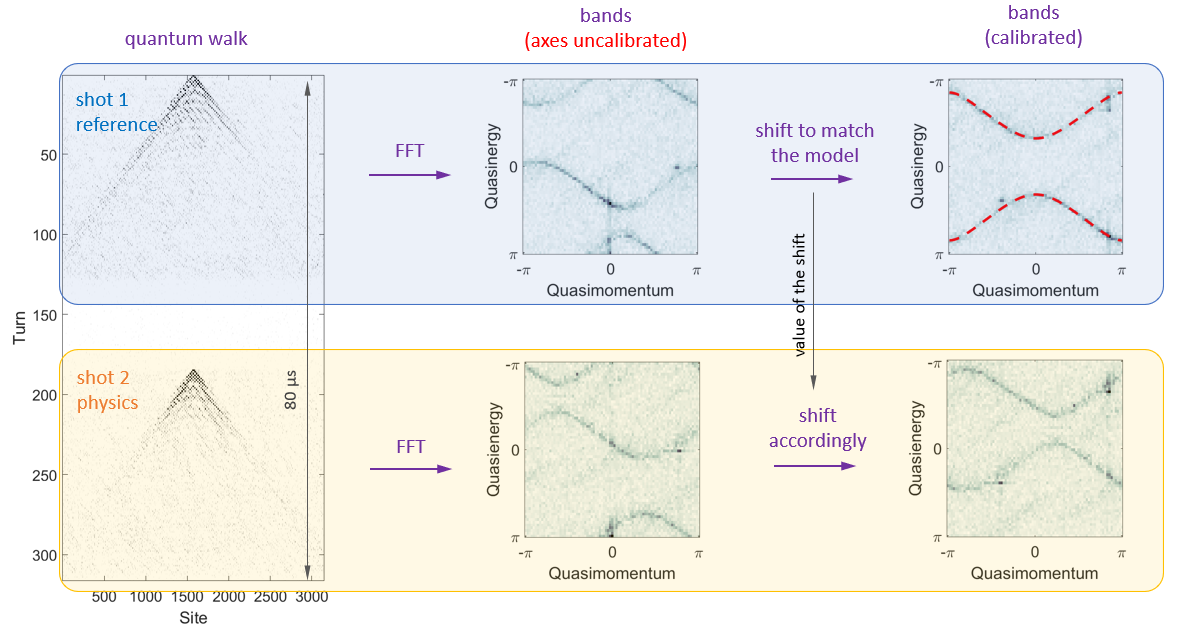}
\caption{\label{figS2} 
Calibration of the band structure.
}
\end{figure}

The length of an optical fiber is sensitive to the environmental temperature and pressure and can fluctuate over time.
For our experiment, this can be thought of as an extra optical phase that the light acquires during its propagation in each fiber ring, which results in a shift of the band structure in both horizontal ($\delta k$) and vertical ($\delta E$) directions.
Over a long time, the length of a fiber ring can change by a few wavelengths.
This implies that both $\delta k$ and $\delta E$ (which are defined modulo $2\pi$) can change in any possible value, and the observed band structure is shifted by an random amount from the expected position.
\\

However, on a short timescale (tens of milliseconds) the length of each ring changes by less than a few percent of a wavelength. 
This allows us to calibrate the band structure by performing two consecutive experimental shots within a short time (100 $\mu$s), during which $\delta k$ and $\delta E$ stay the same.
The first shot implements a simple model without extra phase modulation (i.e. $\varphi=0$), which has a well-known band structure for given $\theta_1$ and $\theta_2$:
\begin{equation}
E_{\pm}^{\mathrm{ref}}(k) =\pm\arccos\left(\cos\theta_{1}\cos\theta_{2}\cos k\right)
\end{equation}
The second shot realizes the experimental system of interest (Fig. \ref{figS2}).
By comparing the band structure of the first shot with its theoretical model we can measure the shifts $\delta k$ and $\delta E$ and therefore calibrate the axes, which will stay the same during the subsequent shot.

%*************************************************************
%*************************************************************
\section{Excitation of a single band}

% To excite a wavepacket in one band we use the technique described in Ref. \cite{wimmer_optical_2013}. 
We start with a theoretical description for the case of $\theta_1=\theta_2=\pi/4$, which has simple and intuitive analytical expressions for the eigenstates.
At time step $m=0$ we inject a train of pulses with Gaussian envelope into one ring, i.e. 
\begin{equation}\label{eq:exc}
\alpha^{0}_n = e^{-\frac{n^2}{\sigma^2}}, \ \beta^{0}_n = 0.
\end{equation}
Such excitation populates the eigenstates with narrow quasimomentum spread around $k\approx0$ in both bands. 
This can be understood knowing that the eigenvectors of the model corresponding to the eigenvalues $E_\pm$ are \cite{wimmer_optical_2013}:
\begin{equation}
\Psi_{\pm} =
\begin{pmatrix}
A \\ B 
\end{pmatrix}_{\pm} = \frac{1}{\sqrt{1+e^{\pm 2\sin k/2}}}
\begin{pmatrix}
1 \\ \mp e^{\pm\sin k/2} e^{-ik/2}
\end{pmatrix} 
\end{equation}
For $k=0$ 
\begin{equation}
\Psi_{\pm} (k=0) = \frac{1}{\sqrt 2}
\begin{pmatrix}
1 \\ \mp 1
\end{pmatrix} ,
\end{equation}
and for broad Gaussian wavepackets with $\sigma\gg 1$ the excitation \eqref{eq:exc} excites equal fraction of both bands at $k=0$:
\begin{equation}
\begin{pmatrix}
\alpha^{0}_n \\ \beta^{0}_n
\end{pmatrix} \approx
\begin{pmatrix}
1 \\ 0 
\end{pmatrix} = \frac{1}{\sqrt 2} \left( \Psi_+ + \Psi_-\right)
\end{equation}
\\

To excite a single band, we program the PM during the turn $m=1$ to apply a phase $\varphi_{1} = \pi/2$. After the first step, the state of the systems becomes
\begin{equation}
\begin{pmatrix}
\alpha^{1}_{n+1} \\ 
\beta^{1}_{n+1}
\end{pmatrix} 
= \frac{1}{\sqrt 2}
\begin{pmatrix}
\alpha^{0}_{n} e^{i\varphi_{1}} \\ 
i\alpha^{0}_{n+2}
\end{pmatrix}
\approx \frac{i}{\sqrt 2}
\begin{pmatrix}
1 \\ 1 
\end{pmatrix} 
= i \Psi_-,
\end{equation}
and, up to the global phase factor, occupies only one single band $\Psi_-$.
Note that choosing $\varphi_{1} = -\pi/2$ would occupy the $\Psi_+$ band.
\\

\begin{figure}[h]
\includegraphics[width=8cm]{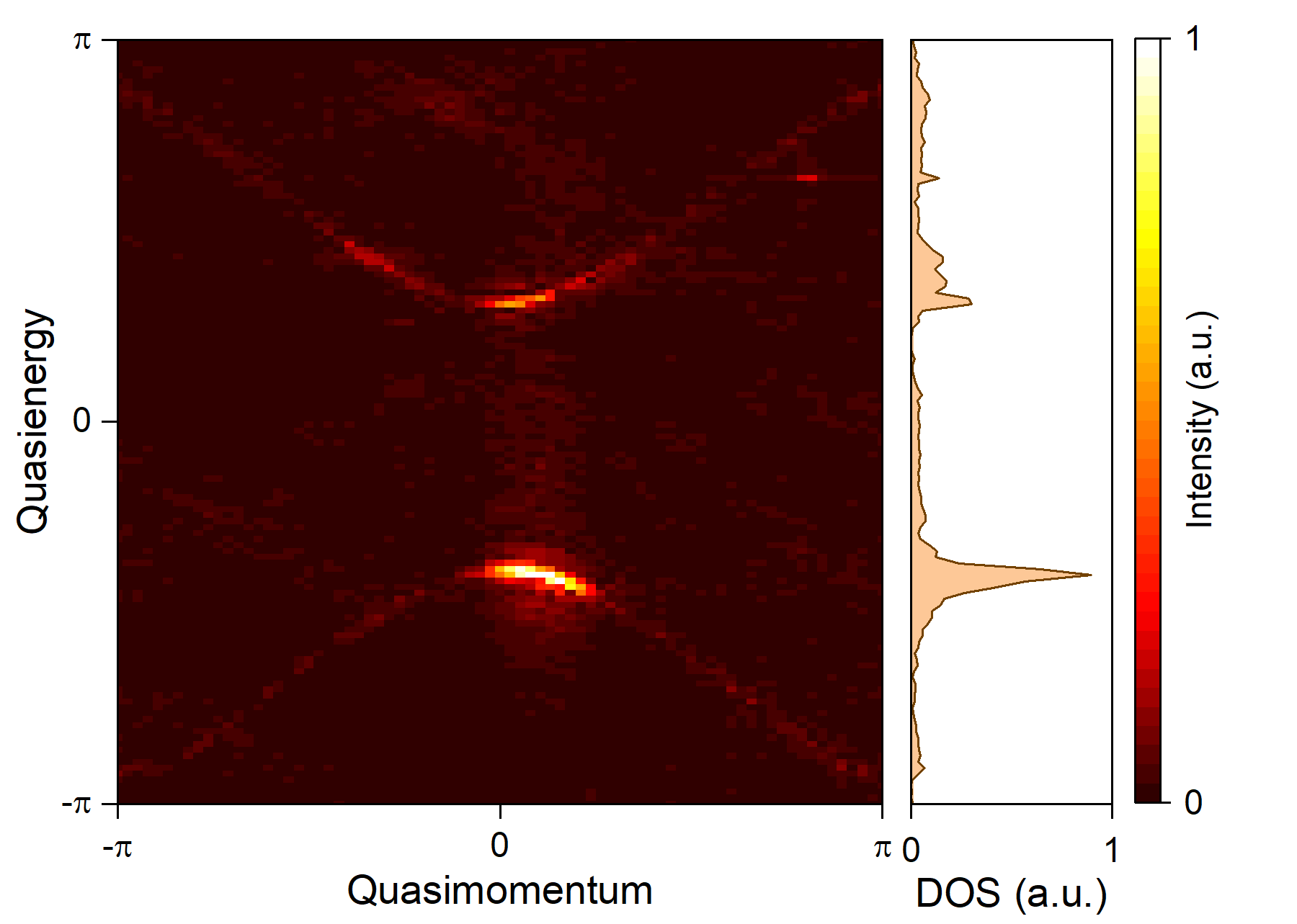}
\caption{\label{figS3} 
Excitation of one band.
}
\end{figure}

For the arbitrary choice of $\theta_1$ and $\theta_2$ one would need to adjust both the phase and the amplitude of signals in two rings in order excite one single band.
However, if chosen values of $\theta_1$ and $\theta_2$ do not alter significantly the shape of the bands (which is the case of our work), one can still transfer the most part of the signal into one band.
In our experiment, we can reproducibly inject more than a 80\% of the emission into one of the bands (Fig. \ref{figS3}).
% Presence of the remaining excitation in the other band manifests itself as a weak signal in the real-space dynamics (Fig. \ref{fig3} of the main text), which does not hinder any of the observed features.

%*************************************************************
%*************************************************************
\section{Identification of trivial and anomalous Floquet phases}

\begin{figure}[h]
\includegraphics[width=12.9cm]{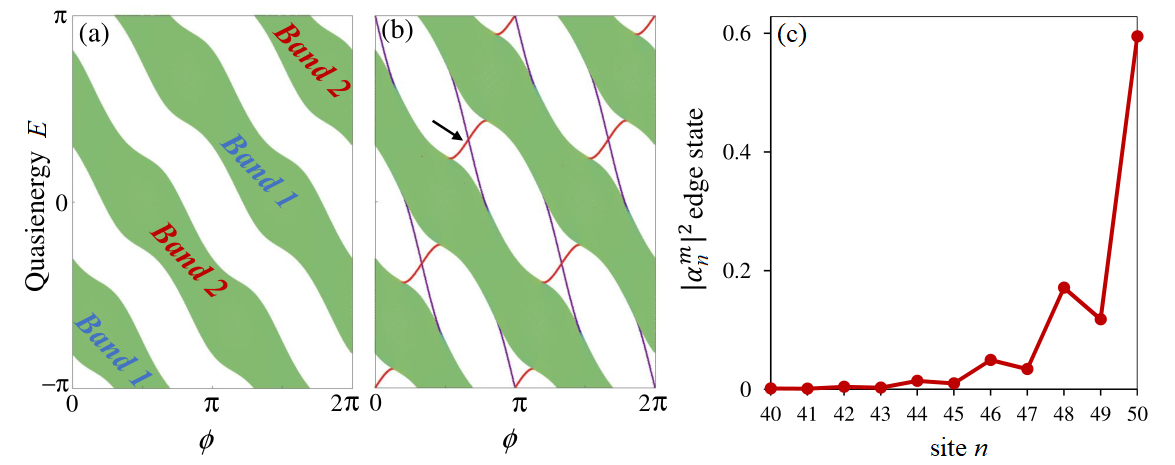}
\caption{\label{figS4} 
Calculated bands for the (a) trivial winding metal with $K=-1$, $\theta_1=\pi/4-0.6$, and $\theta_2=\pi/4$, 
(b) anomalous winding metal with $K=-1$, $\theta_1=\pi/4$, and $\theta_2=\pi/4-0.6$. 
Lines traversing the gap correspond to states localized at the edges.
Both models are comprised of 50 sites along the synthetic dimension.
(c) Probability amplitude of the red edge state marked by a black arrow in (b) at $\varphi=1.24\pi$.
}
\end{figure}

To identify the trivial and anomalous Floquet phases we compute the quasienergy spectra $E(k,\varphi)$ for a finite size system containing 50 unit cells along the synthetic dimension with fully reflective boundary conditions.
The calculated spectra in the trivial and anomalous case are shown in Fig.~\ref{figS4} A and B respectively.
The anomalous phase clearly shows spectral features traversing the gaps, which correspond to the states localized at the edges of the lattice as in Fig. ~\ref{figS4}C.

\end{document}